\documentclass [aps,prd,reprint,eqsecnum,nofootinbib]  {revtex4-1}
\usepackage{amssymb}
\usepackage{amsmath}
\usepackage{bm}% bold math
\usepackage{graphicx}% Include figure files

\def\A{\mathsf{A}}
\def\B{\mathsf{B}}
\def\F{\mathsf{F}}
\def\K{\mathsf{K}}
\def\J{\mathsf{J}}
\def\D{\mathsf{D}}

\def\SO{\mathsf{SO}}
\def\so{\mathsf{so}}

\def\h{\mathsf{h}}

\def\SO{\sf{SO}}

\begin{document}

\title{%
A note on gravity, entropy, and BF topological field theory}
\author{Jerzy Kowalski-Glikman}
\email{jkowalskiglikman@ift.uni.wroc.pl} \affiliation{Institute for
Theoretical Physics, University of Wroc\l{}aw, Pl.\ Maxa Borna 9,
Pl--50-204 Wroc\l{}aw, Poland}

\date{\today}
\begin{abstract}
In this note I argue that the expression for entropic force, used as a starting point in Verlinde's derivation of Newton's law \cite{Verlinde:2010hp}, can be deduced from first principles if one assumes that that the microscopic theory behind his construction is the topological $\SO(4,1)$ BF theory coupled to particles.
\end{abstract}

\maketitle

\section{Introduction}

There is a number of evidences indicating a deep relation between gravity and thermodynamics. In early 1970s four laws of black hole dynamics have been formulated \cite{Bardeen:1973gs}, whose form closely resembled the four laws of thermodynamics. It was then realized that this similarity between gravity and thermodynamics reaches far beyond formal analogy: the bold conjecture of Bekenstein \cite{Bekenstein:1973ur}
that area of black hole horizon is proportional to thermodynamical entropy has been strengthened by Hawking discovery of black hole radiation \cite{Hawking:1974sw}. It turned out that indeed, as suggested by four laws of black hole dynamics, black holes behave as a thermal systems, with entropy and temperature proportional to the area and surface gravity, respectively.

About twenty years later, in a remarkable paper Jacobson \cite{Jacobson:1995ab} 
 has shown that from the proportionality between area and entropy \cite{Bekenstein:1973ur} taken as a fundamental principle one can derive the full Einstein equations of gravity. This idea has been then discussed in depth by  Padmanabhan and others; see \cite{Padmanabhan:2009vy} for recent review and references.

 Building on these developments, in a recent paper \cite{Verlinde:2010hp}, Erik Verlinde  argued that the force of the second law of dynamics and that of Newton's law of gravity can both have their origin in thermodynamics, and can be understood in terms of the entropic force (similar idea, based on equipartition of energy, has appeared earlier in \cite{Padmanabhan:2009kr}.)  Within weeks several follow-up works appeared, testing this idea in various contexts (see, for example \cite{Li:2010cj} for the discussion in the context of cosmology and \cite{Wang:2010px}
 for derivation of the Coulomb law from thermodynamics.) In particular, in \cite{Smolin:2010kk} Smolin argued that Verlinde's proposal can be naturally realized in the context of Loop Quantum Gravity, and suggested its relations with constrained topological field theories. This idea is the starting point of the present work.

 Let us recall the major points of Verlinde's reasoning. The  basic postulate of his work (see \cite{Verlinde:2010hp} for detailed discussion) is the following assumption

 \begin{itemize}
 \item Consider a holographic screen $\cal S$. If particle of mass $m$  crosses the screen, than the change of entropy of the screen is proportional to mass and displacement~$\Delta x$
     \begin{equation}\label{1}
        \Delta S \sim m \Delta x\, .
     \end{equation}
 \item It then follows from the first law of thermodynamics that if there is the temperature $T$ that can be associated with the screen, then there exists the entropic force $F$ satisfying
     \begin{equation}\label{2}
        F \Delta x=T\Delta S\, ,
     \end{equation}
     so that
     \begin{equation}\label{3}
        F \sim mT\, .
     \end{equation}
\end{itemize}

As shown by Verlinde the Newton's law of gravity can be  derived assuming just from this postulate, energy equipartition, and the holographic principle. The reasoning goes as follows. Consider a spherical screen $\cal S$ at the center of which a localized, static chunk of matter of mass $M$ is placed. Assume that the radius of the screen is much larger than the size of the chunk, so that we can assume spherical symmetry of the problem. The holographic principle says that the number of bits $N$ on the screen $\cal S$ is proportional to its area\footnote{In what follows I will use the units in which the velocity of light $c$, the Planck constant $\hbar$, and the Boltzmann constant $k_B$ are all equal 1.}
\begin{equation}\label{4}
    N = \frac{A}{G} \, ,
\end{equation}
which is essentially the statement that the screen $\cal S$ is made of pixels of Planck size (this is the point where, as pointed out in \cite{Smolin:2010kk}, Loop Quantum Gravity with its quantization of area operator \cite{Rovelli:1994ge} naturally fits.) To complete the derivation of Newton's law one has to assume the equipartition of energy on the screen, from which it follows the relation between energy $E=M$ and the temperature
\begin{equation}\label{5}
    M=\frac12\, NT\, .
\end{equation}
Finally, assume that the area of the screen $\cal S$ is
\begin{equation}\label{6}
    A = 4\pi R^2\, .
\end{equation}
From equations (\ref{4}--\ref{6}) it follows that the temperature satisfies
\begin{equation}\label{7}
    T = \frac{2GM}{4\pi R^2}\, ,
\end{equation}
which, together with the postulate (\ref{3}) reproduces the Newton's law, $F=GMm/R^2$. It is worth noticing that for Schwarzschild black hole horizon $R=2GM$, eqn.\ (\ref{7}) reproduces the correct expression for Bekenstein--Hawking temperature $T_{BH}=(8\pi GM)^{-1}$.

As argued by Verlinde the above reasoning is robust and general, the only weak point being the origin of the entropic force (\ref{2}), (\ref{3}). Certainly, there must be some microscopic degrees of freedom responsible for its emergence, and below I will argue that they, and the corresponding force, arise quite naturally in the  formulation of gravity as a constrained topological BF theory.

The plan of this note is as follows. In the next section I will recall the formulation of gravity as a constrained $\SO(4,1)$ BF theory and its coupling to particles. These technical results will be needed for the derivation of Verlinde's entropic force. The reader might decide to skip these technicalities and jump directly to Section III, where the main argument of the paper will be presented. The last section is devoted to discussion and conclusions.

\section{Gravity as a constrained topological field theory}

It is well known for quite some time that gravity can be formulated as a constrained topological field theory. The most popular popular model of this kind is given by Plebanski action \cite{Plebanski:1977zz}, being an action of the constrained BF theory of Lorentz $\SO(3,1)$ group. This model is a starting point for four dimensional spin foam models building (see e.g., \cite{Perez:2003vx}.) In the present context, however, it will be convenient to consider a different model, based on de Sitter gauge group $\SO(4,1)$ (the anti de Sitter model can be constructed analogously.) The main reason for this choice is that the $\SO(4,1)$  model allows for natural particles coupling.

The action of the $\SO(4,1)$ constrained BF theory has the following form \cite{Smolin:2003qu},  \cite{Freidel:2005ak}
\begin{equation}\label{8}
S=\int \B^{IJ} \wedge \F_{IJ} -\frac{\beta}{2} \B^{IJ} \wedge \B_{IJ}  -\frac{\alpha}{4} \B^{ab} \wedge \B^{cd} \epsilon_{abcd}\, ,
\end{equation}
where $\F_{IJ}$ is the curvature of the $\SO(4,1)$ connection one-form $\A_{IJ}$ and $\B^{IJ}$ is a two-form valued in  the algebra  of the $\SO(4,1)$ group. Here the algebra indices $I, J, \ldots$ take values $0, \ldots, 4$, wile the indices $a,b,\ldots=0,\ldots, 3$  label Lorentz subalgebra $\SO(3,1)$ of $\SO(4,1)$. If one decomposes the connection $\A_{IJ}$ into Lorentz and translational parts
\begin{equation}\label{9}
    \A^{ab}=\omega^{ab}, \quad \A^{a4} = \frac1\ell\, e^a \, ,
\end{equation}
solves the equations of motion for $\B^{IJ}$ resulting from (\ref{8}) and plugs the result back to this action, as a result one gets the first order action of General Relativity 
 $$
 S= \frac{1}{2G}\int
 R^{ij}(\omega)\wedge e^k \wedge e^l \epsilon_{ijkl} -
\frac{\Lambda}{12G}\int e^i \wedge e^j \wedge e^k \wedge e^l
\epsilon_{ijkl}
 $$
 accompanied by Holst term and a number of topological  terms (see \cite{Freidel:2005ak} for details.) To get the action of General relativity the coupling constants $\alpha$, $\beta$ of the action (\ref{8}) and the length scale $\ell$ necessary for making the tetrad $e_\mu^a$ dimensionless are to be related to Newton's constant $G$, cosmological constant $\Lambda$, and Immirzi parameter $\gamma$ as follows
\begin{equation}\label{10}
    \gamma=\frac{\beta}{\alpha}, \qquad \frac{1}{\ell^2}=\frac{\Lambda}{3}, \qquad G=\frac{3\alpha(1-\gamma^2)}{\Lambda}
\end{equation}

Let us pause for a moment to comment on the structure of the action (\ref{8}). If  $\alpha$ vanishes the resulting action is just that of a topological field theory with no dynamical degrees of freedom. The local degrees of freedom of gravity (like gravitational waves or the presence of Newton's potential) appear only if the gauge breaking term, controlled by the coupling constant $\alpha$ is nonzero. Therefore the action (\ref{8}) clearly exhibits the split between topological and local degrees of freedom. In other words it is only the last term of (\ref{8}) that knows about dynamics of gravity. In the context of the present paper an obvious question arises: is it possible that the topological action describes the primary degrees of freedom of theory, while the gauge breaking term (i.e., gravity) arises as an emergent phenomenon from entropic force? As it is argued below at least Newton's force between massive bodies can be understood in this way.

It is worth noticing also that the form of the gauge breaking term in the action (\ref{8}) is justified only by the fact that the theory described by this action turns out, at the end of the day, to be equivalent on shell to General Relativity. Therefore, it would be very interesting to find a principle, which would explain the presence of this term (see \cite{Smolin:2003qu} for an interesting proposal in this context.) Deducing Newton's law is the first step in this direction.

As explained in \cite{Freidel:2006hv}, one can straightforwardly add point particles to the theory described by (\ref{8}) by identifying them with Wilson lines. To do that one includes the localized breaking of the gauge symmetry along the one dimensional world-line. The gauge degrees of freedom are
then promoted to dynamical degree of freedom, which, in the case $\alpha\neq0$ reproduce the
dynamics of a relativistic particle coupled to gravity. For a single particle this idea is
realized by choosing a world-line  $\cal P$ and a fixed element
$\K$ in the Cartan subalgebra of the $\so(4,1)$ Lie algebra generated by two generators, the translational $T^{04}$ and rotational $T^{23}$ ones,
depending on the particle rest mass and spin\footnote{Here we consider massive particles only. An extension to the case of massless particles is straightforward.}
\begin{equation}\label{11}
    \K\equiv m \ell\, T^{04} + s
T^{23}
\end{equation}
Note that the particle mass arises quite naturally in this picture in a purely algebraic way and is related to the one of the Casimirs of the gauge algebra.
Then the action for the particle at rest takes the form
\begin{equation}\label{12}
    S_{P}(\A) =-\int
\mathrm{d} \tau\, \mbox{Tr}
 \left(\K \A_\tau(\tau)\right)
\end{equation}
where $\tau$ parametrizes the world line $z^\mu(\tau)$ and
$\A_\tau(\tau) \equiv \A_\mu(z(\tau))\, \dot{z}^\mu$.

The action of the particle moving in an arbitrary way is obtained by realizing that the moving particle is related to the one at rest by an appropriate $\SO(4,1)$ transformation acting on the world line. In this way the gauge degrees of freedom at the location of the particle become its physical degrees of freedom. Thus the Lagrangian of the dynamical particle has the form
\begin{equation}\label{13}
    L(z,\h; \A) = -\mbox{Tr}
\left(\K \A^\h_\tau(\tau)\right)\quad S = \int\, d\tau\, L(z,\h; \A)\, ,
\end{equation}
with
$$
\A^\h=\h^{-1} \A \h + \h^{-1} d \h\, ,
$$
which can be rewritten as
\begin{equation}
L(z,\h; \A)= L_1(z,\h)-\mbox{Tr}(\J \A_\tau) \, , \label{14}
\end{equation}
with the first being the  particle kinetic Lagrangian
 \begin{equation}
L_1(z,\h) =  -\mbox{Tr} (\h^{-1}\dot \h \K)\, , \label{15}
\end{equation}
while the second  describes its coupling to the connection $\A$, with  $\J$ being the dynamical particle momentum/spin and is given by
\begin{equation}\label{16}
    \J \equiv
\h\, {\K} \,\h^{-1}\, .
\end{equation}
It can be shown that from (\ref{14}) the correct particle equation of motion (Mathisson--Papapetrou equation) follows; the theory described by (\ref{14}) and (\ref{8}) leads to Einstein-Cartan equations with point sources carrying mass and spin (see \cite{Freidel:2006hv} for detailed discussion.)

This completes our description of the theory. Let us now turn to the discussion of solutions of topological BF theory coupled to such defined particle.

Take the topological limit $\alpha\rightarrow0$ in (\ref{8}) and (\ref{14}) and consider the resulting field equations \cite{KowalskiGlikman:2006mu} for the particle at rest at the origin of an appropriate coordinate system\footnote{The reader may wonder that using  the coordinates, and the geometry, we let gravity sneak through the back door. Of course, we need geometry to formulate the model, but the relation between local degrees of freedom of gravity and geometrical quantities is not present at this level yet, because dynamical gravity is not there.}  One finds
\begin{equation}\label{17}
    \F^{IJ} =  \beta     \B^{IJ}
\end{equation}
\begin{equation}\label{18}
   \D_\A\, \B^{IJ} = \J^{IJ}\, \delta_P, \quad \delta_P =\delta^3(x) \varepsilon
\end{equation}
where $\D_\A$ is covariant derivative of connection $\A$ and
$\varepsilon$ is the volume three-form on a constant time surface.

If one solves (\ref{17}) for $\B$ and substitutes the result to (\ref{18}) one finds that the left hand side of the resulting equation is zero by virtue of Bianchi identity. It is clear therefore that there does not exist a nonsingular connection $\A$ satisfying these equations for nonzero source. However, if one allows connections with string-like singularity (Misner string \cite{Misner:1963}, which is the gravitational counterpart of Dirac string) these equations can be solved.

In fact it turns out that a pointlike source must be accompanied by a string extending from the source to infinity. As it was argued in \cite{KowalskiGlikman:2006mu} the spacetime corresponding to the solution of these equations\footnote{In the limit $\ell\rightarrow\infty$, which corresponds to the vanishing cosmological constant. Since the Taub-NUT charge $n$ does not depend on $\ell$, taking the limit does not influence it.} is the (linearized) Taub-NUT spacetime

\begin{equation}\label{19}
      g = -\left(dt + n(1-\cos\theta)d\phi\right)^2 + dr^2 + r^2(d\theta^2 + \sin^2\theta\,
   d\phi^2)
\end{equation}
with Taub-NUT charge
 \begin{equation}\label{20}
    n=\bar G m, \quad \bar G = G\, \frac\gamma{1+\gamma^2}\, .
\end{equation}

This completes our brief description of  constrained $\SO(4,1)$ BF theory, its relation to gravity and coupling to point sources.

\section{Entropy and gravity from topological field theory}

In the previous section I argued that if one couples  the $\SO(4,1)$ topological BF theory (which after gauge breaking down to $\SO(3,1)$ is equivalent to General Relativity) to point particles, then the theory forces the particles to be accompanied by semi-infinite Misner strings. Moreover, the space-time corresponding to such solution is  the Taub-NUT solution linearized in the charge $n$, which turns out to be proportional $Gm$, where the particle mass $m$ is the value of one of the Casimirs of $\SO(4,1)$ (the second one describes the spin, but here we discuss the spinless case only.)

Knowing this let us turn to  deducing the form of entropic force acting on the particle. Suppose the test particle of mass $m$ is at distance $R$ from the mass $M$, which we can assume to be also point-like. Consider now, as in Verlinde's argument, a spherical screen $\cal S$ of radius $R$. Let the test particle move radially towards the central mass piercing the screen, and let its displacement be  $\Delta x$. As a result we have now a segment of the Misner string of the test particle of the length $\Delta x$ connecting it with the screen. Therefore the screen that previously was just a sphere\footnote{More precisely $\cal S$ consists of the sphere along with the attached string (or strings) emanating from the central mass $M$. But since we are only interested in the (infinitesimal) change of entropy, we do not have to consider them.} now becomes a sphere with a  piece of Misner string, the line segment of  length $\Delta x$ attached, $\cal S'$.

Let me now turn to the main argument of this paper. It is well known that there is entropy associated with Misner string, see \cite{Hawking:1998jf}, \cite{Hawking:1998ct}, \cite{Carlip:1999cy}, and
\cite{Mann:1999pc} where it is argued that the entropy of Misner string is intrinsically defined. In particular, using methods of conformal field theory Carlip \cite{Carlip:1999cy} shows that the segment of the Misner string of the length $\Delta x$ carries the entropy
\begin{equation}\label{21}
    \Delta S = \frac1{8\pi G}\, n\, \Delta x= \frac1{8\pi }\, m\, \Delta x\, .
\end{equation}
Although this result has not been rigorously established in the present context of BF theory, it is unlikely that a formula analogous to (\ref{21}) does not hold in this case as well. It seems clear that Misner string carries entropy, no matter what is the theory describing local and/or topological degrees of freedom. If one accepts this argument, it follows from simple dimensional analysis that the entropy of the segment of Misner string of the length $\Delta x$ has to have the form
\begin{equation}\label{22}
    \Delta S = \zeta\, m\, \Delta x\, ,
\end{equation}
where $\zeta$ is the coefficient depending on the structure (and coupling constants) of the underlying theory.

The entropy (\ref{22}) adds to the original entropy of the screen, and since it is proportional to the test particle displacement it leads to the emergence of the entropic force. Notice that since entropy increases when the test particle moves towards the mass $M$ this entropic force is attractive. Also when the test particle which was initially inside the screen moves outside, the entropy decreases, since the contribution from the Misner string is no longer present.

Having (\ref{22}) it is possible now to run the remaining part of the Verlinde's argument essentially without modifications. The only point that is worth discussing is the equation relating the number of the screen pixels with area. Why $G$ is the measure of area of a pixel? In Loop Quantum Gravity this question finds its natural answer thanks to the fact that quantization of area in Planck scale units is the main result of this theory. It is not excluded that even in the context of BF theory one can define area operator with discrete spectrum. Until this idea is supported (or disproved) by concrete calculations we can rely only on general intuitions. The theory at hands provides us with the dimensionful scale $\ell$ and the dimensionless coupling constant $\beta$. From the two it is possible to construct another constant of dimension of area
\begin{equation}\label{23}
    \bar G = \frac{3\beta}{\ell^2}
\end{equation}
which in the full theory (including nontrivial gauge breaking term) becomes proportional to Newton's constant of general relativity (cf.\ (\ref{10})). Since $\beta$ has some final value, and since $\ell$ is an infrared scale of the theory, it is quite natural to treat $\bar G$ as an intrinsic ultraviolet scale of the theory, and thus to  replace (\ref{4}) with
\begin{equation}\label{24}
    N = \frac{A}{\bar G}, \quad A=4\pi R^2\, ,
\end{equation}
which directly, by virtue of Verlinde's argument recalled in Introduction leads to the Newton's law
\begin{equation}\label{25}
    F=\frac{G\, m\, M}{R^2} \, ,
\end{equation}
where $G=4\pi \zeta\bar G$ is the Newton's constant, whose value can be directly measured, e.g., in Cavendish experiment. This concludes the presentation of the main argument of this paper.
%\newline

\section{Conclusions and outlook}

In this note I argued that the form of entropic force being the starting form of the recent proposal of Verlinde \cite{Verlinde:2010hp} to seek the origin of gravity in thermodynamics can be understood if one assumes that the fundamental degrees of freedom behind it are described by the topological BF theory coupled to particle(s). The reason for this is that, as shown in \cite{Freidel:2006hv} and discussed in \cite{KowalskiGlikman:2006mu}, a particle carrying the charge of (anti) de Sitter $\SO(4,1)$ ($\SO(3,2)$) group coupled to the topological BF theory with the same gauge group must have Misner string attached. This string, in turn, carries entropy, which adds to the entropy of the holographic screen $\cal S$ when the particle crosses it, which results in emergence of the entropic force.

There are at least two problems that has to be solved before this idea becomes a solid proposal. First, one has to calculate the entropy of Misner string directly in the framework of BF theory, to fix the constant $\zeta$ in (\ref{22}). This can be presumably done with the help of the method similar to that described in \cite{Carlip:1999cy}. Second, it would be interesting to see if it is possible to improve on the part of the original Verlinde's argument that makes use of equipartition of energy to get the expression for temperature (\ref{7}). It would be desirable in this context to check if one can define for BF theory an area operator with discrete spectrum, resembling this of Loop Quantum Gravity \cite{Rovelli:1994ge}. The work on both these problems is in progress.

\begin{acknowledgments}
    I would like to thank Lee Smolin for discussion and encouragement. This work is supported in part by research projects N202 081 32/1844 and NN202318534 and Polish Ministry of Science and Higher Education grant 182/N-QGG/2008/0.
\end{acknowledgments}

\end{document}